# Magnified Image Spatial Spectrum (MISS) microscopy for nanometer and millisecond scale label-free imaging

**Hassaan Majeed**[1#], **Lihong Ma**[2#], **Young Jae Lee**[3], **Mikhail Kandel**[1], **Eunjung Min**[4], **Woonggyu Jung**[5], **Catherine Best-Popescu**[3], **Gabriel Popescu**[1*]

[1] *Quantitative Light Imaging (QLI) Lab, Beckman Institute of Advanced Science and Technology, University of Illinois at Urbana Champaign, 405 N Matthews, Urbana, IL 61801, USA.*
[2] *Institute of Information Optics, Zhejiang Normal University, Jinhua 321004, China.*
[3] *Neuroscience Program, Micro and Nanotechnology Lab, University of Illinois at Urbana Champaign, 208 N Wright St., Urbana, IL 61801, USA.*
[4] *Rowland Institute at Harvard University, 100 Edwin Land Blvd, Cambridge, MA 02142, USA.*
[5] *Department of Biomedical Engineering, Ulsan National Institute of Science and Technology, 50 UNIST-gil, Ulsan 44919, Republic of Korea.*
[#] *These authors contributed equally.*
*gpopescu@illinois.edu.

**Abstract:** Label-free imaging of rapidly moving, sub-diffraction sized structures has important applications in both biology and material science, as it removes the limitations associated with fluorescence tagging. However, unlabeled nanoscale particles in suspension are difficult to image due to their transparency and fast Brownian motion. Here we describe a novel interferometric imaging technique referred to as Magnified Image Spatial Spectrum (MISS) microscopy, which overcomes these challenges. The MISS microscope provides quantitative phase information and enables dynamic light scattering investigations with an overall optical path length sensitivity of 0.95 nm at 833 frames per second acquisition rate. Using spatiotemporal filtering, we find that the sensitivity can be further pushed down to $10^{-3}$-$10^{-2}$ nm. We demonstrate the instrument's capability through colloidal nanoparticle sizing down to 20 nm diameter and measurements of live neuron membrane dynamics. MISS microscopy is implemented as an upgrade module to an existing microscope, which converts it into a powerful light scattering instrument. Thus, we anticipate that MISS will be adopted broadly for both material and life sciences applications.

**OCIS codes:** (180.0180) label-free microscopy; (290.0290) light scattering; (110.0110) millisecond imaging; (110.0110) nanoscale imaging; (110.3175) off-axis interferometry.

## 1. Introduction

Unlabeled nanoscale objects, such as colloidal nanoparticles and subcellular vesicles, are extremely challenging to image with light for several reasons [1]. Transparent samples only modulate the phase of the incident field, which makes them undetectable by intensity-based imaging systems. While systems that detect the scattered field are able to register signals from nanoparticles [2], they are unable to detect all spatial frequencies in a single acquisition which requires measurement of the phase image. Furthermore, objects of small spatial scales tend to move fast, adding constraints to the throughput of the imaging system. Here we introduce magnified image spatial spectrum (MISS) microscopy, which meets these challenges. We show that because of the particular interferometric geometry used, MISS microscopy is highly sensitive to phase or optical path-length difference (OPD) changes introduced by nanoscale objects and can achieve high acquisition rates. Therefore, our measurements can quantify the dynamics of particles that are much smaller than the wavelength of light.

The MISS system is a quantitative phase imaging (QPI) [3] instrument that combines the benefits of speed, specific to *off-axis holography* [4], and phase sensitivity, associated with *common-path interferometry* [5]. QPI is a rapidly growing biomedical research field [6]. In QPI, the image consists of the OPD introduced by the specimen at each point in the field of view. This quantity reports on the product of the refractive index and thickness [7] of a specimen, also allowing extraction of its dry mass density [8-10]. The refractive index and thickness information can be decoupled by recording QPI data along another dimension, e.g., angle [11, 12], spectrum [13], z-axis [14, 15] and, thus, allowing tomographic imaging of the object. As a label-free approach, QPI eliminates photobleaching [16] and reduces phototoxicity [17] that often prove challenging in fluorescence microscopy. This capability has led to imaging applications in cellular biology [10, 18-28], disease diagnosis [29-36] as well as in material science [5, 37-39].

MISS microscopy provides complete information about the complex imaging field. As a result, this wave can be *numerically* propagated to any arbitrary plane, including the *far-zone*, where typical angular scattering measurements are performed. From a single MISS image we obtain angular scattering information simultaneously at all angles within the numerical aperture of the objective. In essence, by adding the MISS module to an existing microscope, we upgrade the existing instrument to the full capabilities of dynamic light scattering [40]. Due to its high acquisition rates, we show that time-resolved MISS imaging data can be used to **(a)** accurately size colloidal particles under Brownian motion, of radii down to 1/50 of the wavelength of light and **(b)** measure membrane dynamics from live, electrically active mammalian neurons, without the need for labels or physical contact.

## 2. Results and Discussion

### 2.1 MISS optical setup

The MISS optical setup is illustrated in Fig. 1. The image plane outside of a commercial microscope (Zeiss Axio Observer Z1) is relayed to the sCMOS camera (Zyla, Andor) plane using a 4f system comprising lenses $L_1$ and $L_2$. A diffraction grating (Edmund Optics, 110 lp/mm) is placed precisely at the image plane outside the microscope to separate the image field into multiple orders, each containing complete information about the object. At the Fourier plane of $L_1$, the zeroth diffraction order is passed through a gradient index (GRIN) lens $L_3$ (Edmund Optics) while the first diffraction order goes through unaffected. In combination with $L_2$, lens $L_3$ forms a 4f system that produces at the camera plane a highly magnified image of the zeroth-order Fourier spectrum. The key component in this 4f system is the microlens $L_3$ which has a very short focal length ($f_3 = 0.3 mm$), leading to a magnification factor of $M = f_2/f_3 = 500$ ($f_2 = 150\,mm$). Since the diameter of the beam entering lens $L_1$ is approximately $d = 2.6\,mm$ (controlled by using an iris at the output port of the microscope), the width of the DC component at the back focal plane of $L_1$ is $1.22\lambda f_1/d = 15.6\,\mu m$, with $\lambda = 0.532\,\mu m$ and $f_1 = 60\,mm$ [41]. After the 500 fold magnification, this spot spans 7.8 mm at the camera plane. Since the width of a 1000 x 1000 pixel image at the camera plane is around 6.5 mm, the camera sensor only detects the DC component of the zeroth order, which provides a uniform, plane wave reference. The sensor essentially performs low-pass filtering on the image-field carried by the zeroth-order. The magnified Fourier transform interferes with the image field from the first order to form an interferogram, which can then be processed to obtain the phase image, $\phi(x, y)$, using a Hilbert transform reconstruction algorithm [42, 43]. Note that the $L_2$-$L_3$ 4f-system is not perfect, in the sense that the back focal plane of $L_3$ and front focal plane of $L_2$ do not overlap perfectly. However, this distance error is only $2f_3 = 0.6$ mm, which is negligible compared to $f_2 = 150\,mm$.

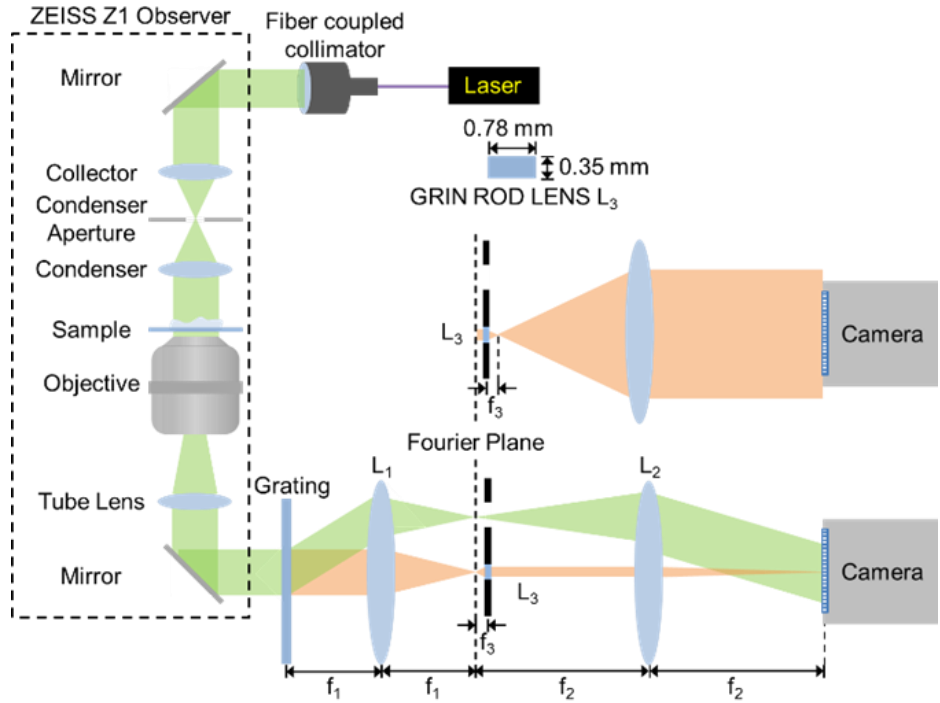

Fig. 1. MISS microscope. $f_1 = 60$ mm, $f_2 = 150$ mm, $f_3 = 0.3$ mm. $L_3$ and $L_2$ form a 4-f system that magnifies the Fourier transform of the zero-order by a factor of 500. For the $L_3$-$L_2$ imaging system, image formation at both high and low spatial frequencies is shown to illustrate the magnification process.

Like MISS, diffraction phase microscopy (DPM) [44-46] also uses a diffraction grating to create a common-path, Mach-Zehnder type interferometer. However, MISS eliminates the need for a micrometer-size pinhole to perform the low-pass filtering and, instead, greatly magnifies the spatial spectrum plane, such that the camera itself plays the role of the pinhole. As a result, the MISS system is significantly more robust and easier to align than DPM, as a 5-10 μm diameter pinhole is now replaced by a 350 μm diameter microlens. With the DPM system, a researcher needs to align the pinhole every time they begin an experiment which is difficult for a 5 μm pinhole. With MISS, typically no adjustments are required. Due to the large diameter of the GRIN lens, MISS is also less susceptible to drift over time, making experiments at long timescales easier. The collection of the reference field is also more efficient in MISS than in DPM, evidence for which has been included in Appendix A. As demonstrated in Fig. 7 (Appendix A), the interferometric visibility in MISS is approximately 2 times that of an equivalent DPM system. Since the phase sensitivity of an off-axis interferometry system is related to the interferometric visibility [47], the MISS microscope provides the same phase sensitivity as DPM but at higher acquisition rates [see Section 2.2].

Throughout our experiments, we used a monochromatic, spatially single mode laser ($\lambda = 532$ nm) as the illumination source. The use of a monochromatic source ensures that the sample and reference waves stay within the coherence length in spite of the latter passing through the GRIN lens. All imaging was carried out using a 40x/ 0.75 NA bright field objective. The accuracy of the MISS microscope was assessed by imaging calibration samples of known refractive index and thickness. We used micro-fabricated quartz pillars having a refractive index of $n = 1.547$ at the illumination wavelength [48] and a height of approximately 80 nm, as measured using an Alpha-step IQ Profilometer. Figure 2 (a) shows the measured interferogram and Fig. 2 (b) its power spectrum. As the spectrum shows, the interferogram generated using the GRIN lens provides modulation at high contrast, evidenced by the relative intensities of the two non-central spectral lobes. Furthermore, the lobes are clearly separated indicating that sampling requirements, which are crucial in off-axis interferometers, have been adequately met [5]. Figure 2 (c) shows the height map of a 40 µm wide square pillar generated from the processed phase image $\phi(x, y)$ as $h(x, y) = \lambda \phi(x, y) / 2\pi(n-1)$. Before computation of the height map, the phase image was spatially low-pass filtered with a kernel size equal to 7 x 7 pixels (half of the width of the system point spread function) in order to remove noise at spatial frequencies above the objective numerical aperture. Figure 2 (d) shows the histogram of pillar height map in Figure 2 (c). The separation of the means of the two peaks was measured to be 80.6 nm which agrees well with the pillar height reported by the profilometer, indicating that the system is able to accurately quantify nanoscale topography. The noise in the pillar height map are largely due to the imperfections in the etching process and sample wear and tear.

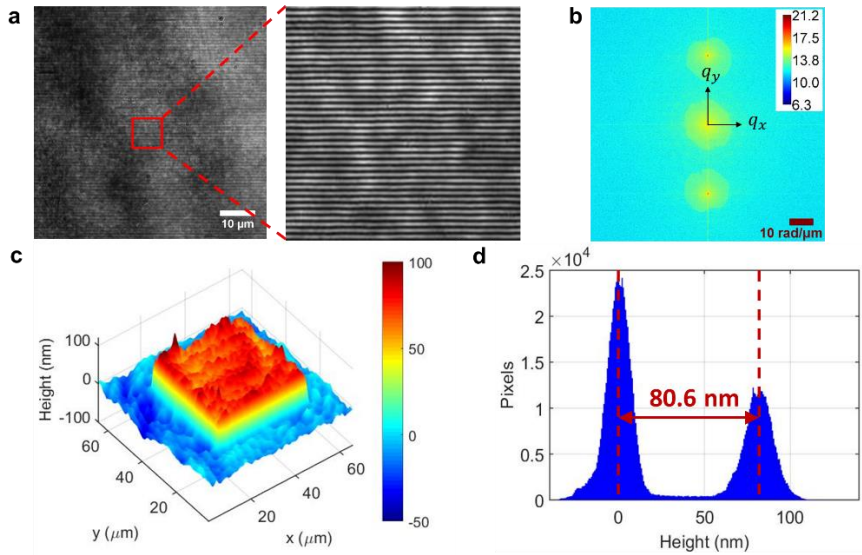

Fig. 2. (a) A 1024 x 1024 raw image of 40 µm wide square quartz pillar obtained using the MISS system (b) The Fourier transform of the raw image. Color bar is in log base 10 arbitrary units. (c) Height map obtained from the reconstructed phase map with color bar in nm. (d) Histogram of the pillar height map. The separation between the peak means corresponds to the measured pillar height.

## 2.2 MISS phase sensitivity

One other key figure of merit of the MISS system is its spatio-temporal noise, which determines the sensitivity of the system or the smallest change in OPD that it can resolve in both space and time. To characterize the noise levels of our system, a stack of 1024 no-sample images (256x1500 pixels) were acquired at 833 fps using an exposure time of 1.2 ms. Figure 3 (a) shows a snapshot of these data. Images of this aspect ratio allowed us to maximize the acquisition rate obtainable from the sCMOS camera. After extracting the phase maps, the stack of images were spatially low pass filtered with a kernel of 7 x 7 pixels (corresponding to half the size of the system point spread function width) to remove noise at spatial frequencies outside the objective numerical aperture. To remove any constant aberration or tilt, a background image was recorded and subtracted from the stack. Figure 3 (b) shows the OPD histogram for the entire stack of no-sample images, which yields the noise standard deviation of $s = 0.95$ nm. This value represents the overall noise performance and includes all the spatio-temporal frequencies captured by our measurement.

The phase sensitivity in off-axis interferometry depends both on the signal efficiency and the exposure time (or frame rate) [3, 47]. Because MISS is more signal efficient than DPM (see Appendix A), it leads to better spatio-temporal phase sensitivity for a given acquisition rate. In previous works, approximately the same phase sensitivity as demonstrated for MISS was obtained for DPM but at lower acquisition rates (97.1 fps in [46] and 8.93 fps in [49]). While Poorya et al. [47] reported 0.28 nm phase sensitivity using DPM, this was also at a lower acquisition speed and they used a specialized camera with a well-depth of 60000 electrons (our camera has a typical depth of 30000 electrons).

To understand how noise affects our measurements at various spatial and temporal scales, we performed spectral analysis. Figure 3 (c) shows the spectral decomposition of the noise variance plotted against spatial frequencies $q_x$, $q_y$ and temporal frequency $\omega$. This decomposition was obtained by taking the Fourier transform of the no-sample stack along all three dimensions ($x, y$ and $t$) and normalizing the result by the total OPD noise variance, $s^2 = 0.9\ nm^2$ (see Section 8.5.3 in [3]). The spectrum in Fig. 3 (c) indicates that if the signal of interest lies within certain spatio-temporal frequency bands, filtering can be used to significantly improve the signal to noise ratio (SNR) of the measurement. Figure 3 (d) illustrates this point by showing the noise level in 4 different frequency bands of interest. Depending on the band chosen, the overall noise can be reduced by 2-3 orders of magnitude through this filtering procedure, pushing the sensitivity limits of the MISS microscope to extremely low values, down to 0.005 nm.

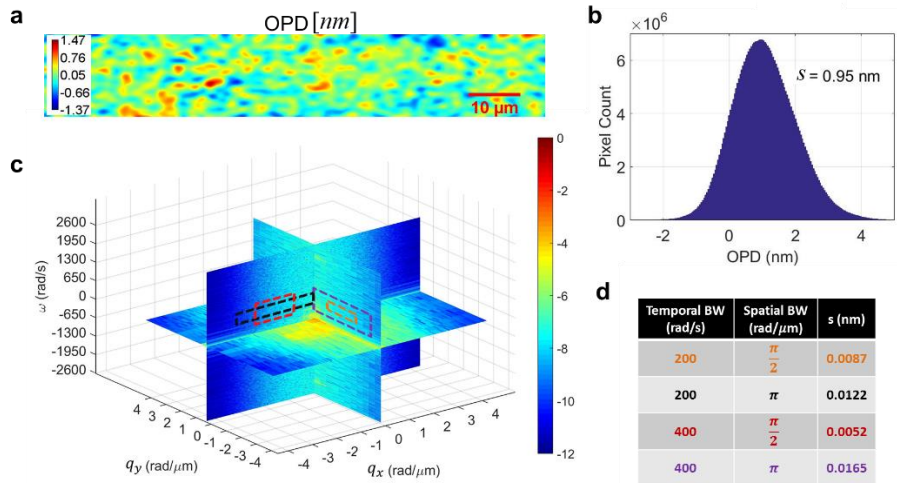

Fig. 3. Analysis of spatiotemporal stability of the MISS microscopy system. (a) A 256 x 1500 pixels no-sample OPD image with color bar in nm. (b) The histogram of the noise OPD stack acquired at 833 fps. (c) Plot showing the noise content at each spatial and temporal frequency component along three different planes in 3d frequency space. Color bar is in log scale with units of $\frac{nm^2}{(rad/\mu m)^2 (rad/s)}$. (d) Band-pass filtering over the spatio-temporal bands shown in (c) results in noise values 2-3 orders of magnitude less than the total noise of 0.95 nm.

### 2.3 Colloidal nanoparticle sizing using MISS

Next, we show that due to the quantitative phase information it provides, MISS can operate as a highly sensitive dynamic light scattering instrument. Thus the instrument can retrieve the particle size of Brownian particles, which are too small to be resolved by a conventional microscope. In traditional dynamic light scattering [50], the intensity of the light scattered by, say, colloidal particles, is measured at a particular angle as a function of time. The temporal autocorrelation of the intensity fluctuations is computed and its decay time can then be related to the translational diffusion coefficient $D$. The diffusion coefficient is defined by the Stokes-Einstein equation [51] as

$$D = \frac{k_B T}{6\pi\eta r} \qquad (1)$$

where $k_B$ is the Boltzmann constant, $T$ the thermodynamic temperature, $r$ the particle radius, and $\eta$ is the dynamic viscosity of the fluid [52]. Because the measurement in our microscope takes place at the image plane, where plane waves from different angles overlap, time-lapse MISS imaging is equivalent to a dynamic light scattering measurement simultaneously at all scattering angles within the objective pupil. Unlike traditional dynamic light scattering, we obtain information from an entire range of spatial frequencies, which allows us to compute a dispersion relation [53, 54]. This relation connects the correlation decay rate (temporal bandwidth) $\Gamma$ and the spatial frequency $q$ (see Section 4: Materials and Methods). For suspended particles, $\Gamma$ is related to the advection velocity distribution width $\Delta v$ and diffusion coefficient $D$ through a second-order polynomial in $q$ [53]:

$$\Gamma = \Delta v q + D q^2. \qquad (2)$$

For colloidal particles undergoing Brownian motion, diffusive motion dominates over advection ($\Delta v \approx 0$) and the dispersion relation is

$$\Gamma = D q^2. \qquad (3)$$

Thus, computing $\Gamma(q)$ from the MISS data yields $D$ which in turn gives access to the particle diameter $p = 2r$, via Eq. (1). Instead of fitting the power spectra at each spatial frequency $q$, which is computationally expensive, we calculate $\Gamma$ in the time domain, using the relationship between moments in the frequency domain and the derivatives in time domain (see Section 4: Materials and Methods for details). This Fourier analysis procedure is referred to as dispersion-relation spectroscopy (DPS) [55].

In order to demonstrate the particle sizing capability of our system, we performed experiments on polystyrene beads (Polysciences Inc., diameters ranging from 1 μm down to 50 nm), as well as 20 nm gold nanoparticles (Sigma Aldrich), all suspended in water at room temperature. While the 20 nm gold nanoparticles are not transparent, determining their size using optical microscopy is still challenging due their small absorption and scattering cross-sections. The samples in each case were prepared for imaging by pipetting a 1 μl drop of nanoparticle solution and sandwiching it between two coverslips. Figure 4 illustrates the approach using analysis results for 1 μm particles. The phase map extracted by our system for a solution of 1 μm particles is shown in Figure 4 (a). Figure 4 (b) shows the map of the temporal standard deviation associated with the OPD. The high values in this map indicate areas of strong phase fluctuations due to particle motion. At the same time, this standard deviation map reveals the trajectories of individual particles. The temporal bandwidth map $\Gamma(q_x, q_y)$, extracted using DPS where $\mathbf{q} = (q_x, q_y)$ represents the wave vector, is shown in Fig. 4 (c). Finally, Fig. 4 (d) shows the profile $\Gamma(q)$ obtained by radially averaging the map in Fig. 4 (c) and the corresponding fit with the quadratic dispersion relation [Eq. (3)]. Because there is always residual space-independent temporal noise in our data, we fit our plots with $Dq^2$ offset by a constant $a$. The results of the fit in terms of the diffusion coefficient $D$ and this background constant are shown in the legend.

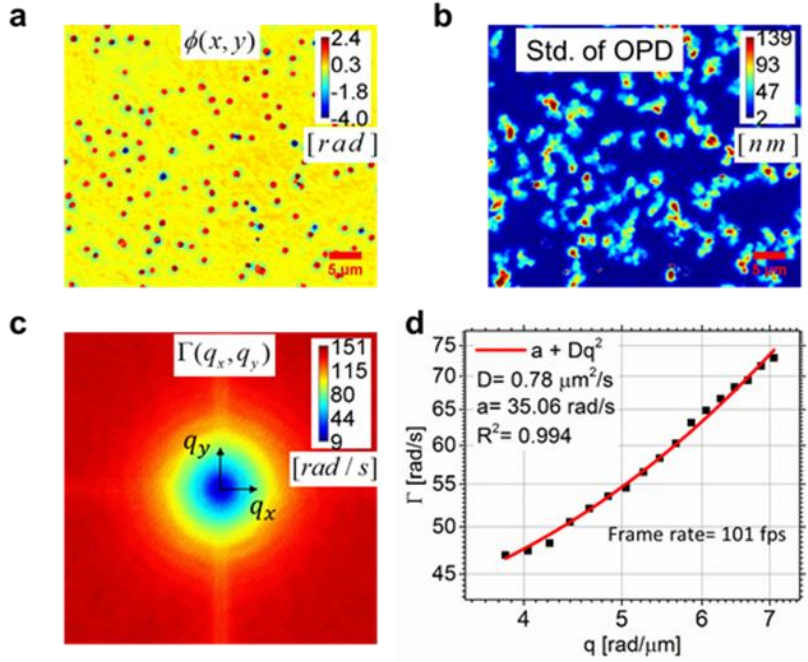

Fig. 4. Particle sizing results for 1 $\mu$m polystyrene particles using dispersion relation spectroscopy (DPS). (a) A representative phase image from the 512 frame stack (color bar in radians). (b) The temporal standard deviation map of OPD computed for the stack (color bar in nm) (c) 2d plot of temporal bandwidth $\Gamma$ versus wave vector $\mathbf{q} = (q_x, q_y)$. (d) Profile obtained by radial averaging of 2d plot in (c). Fitting a second order polynomial to the curve allows the determination of the diffusion coefficient $D$, as indicated.

Figure 5 summarizes the analysis for additional 4 particle sizes, with one representative experiment each. For each particle size, we repeated the experiment 20 times. The diffusion coefficients $D$ obtained from these experiments are presented in Figure 5 (e) in terms of average and standard deviation error. The linear relationship between $\log_{10}(D)$ and $\log_{10}(p)$ yields a slope of ~ - 0.99, which agrees with the Stokes-Einstein relation expressed in Eq. (1). According to this equation, the y-intercept of this straight line should theoretically equal $\log_{10}(k_B T / 3\pi\eta) =$ 2.74 at $T = 293K$ and $\eta = 7.9$ x 10-4 Pa.s, which is comparable to the experimental value of 2.81. The slight discrepancy is likely due to small errors in the assumed values of viscosity and temperature.

Note that MISS yields accurate results of particle diameters even when these values are significantly below the diffraction limit, which is on the order of $\lambda / NA \approx 700\,nm$. In essence, this achievement is possible because information about the particle size is not incorporated in the scattering angle (momentum) distribution, but rather in the temporal fluctuations of the OPD. As a result, in principle, there is no physical bound to how small a particle MISS can detect. This limit is governed in practice by two factors: 1) the OPD sensitivity of the instrument, which becomes challenging as the particles under investigation decrease in size, and 2) the acquisition rate, which has to increase with decreasing particle size. MISS can correctly quantify phase fluctuations from 20 nm gold nanoparticles because of its particular interferometric geometry that allows for fast and sensitive measurements.

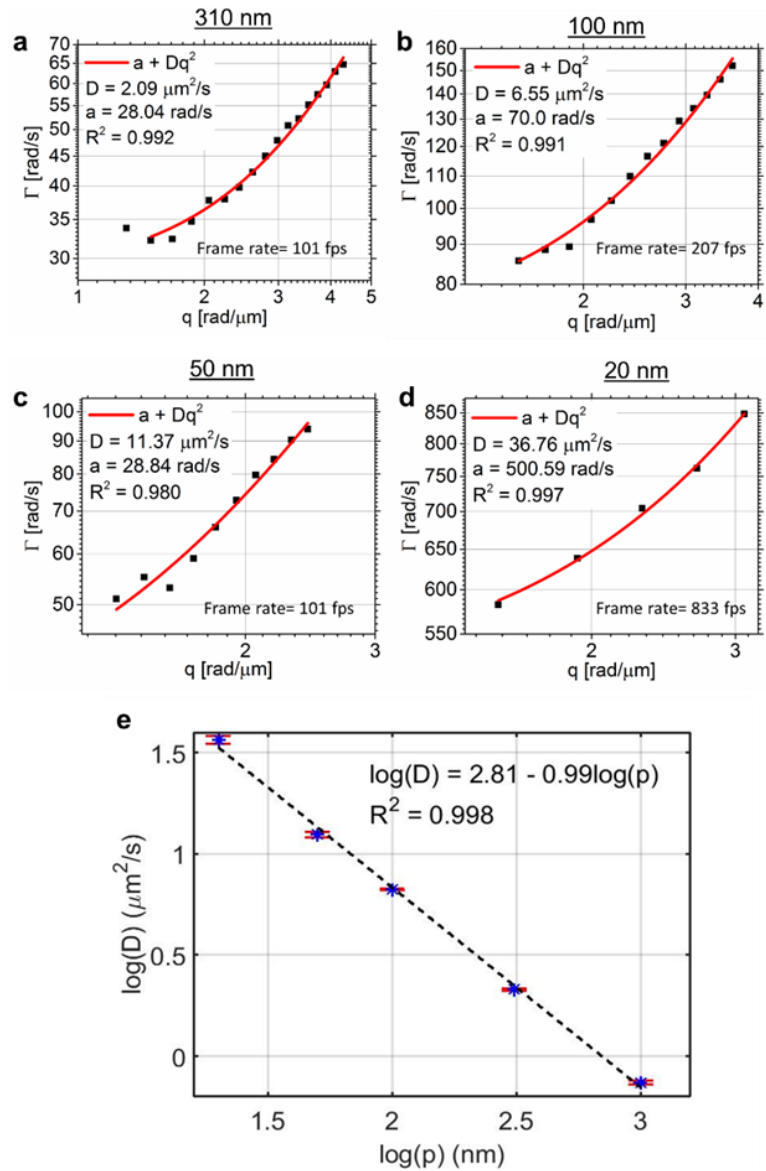

Fig. 5. (a, b, c, d) DPS analysis results for 4 different particle sizes in water, as indicated. (e) Fitting a first-order polynomial to the $\log_{10}(D)$ vs $\log_{10}(p=2r)$ data verifies that the measurement follows the Stokes-Einstein equation. Each data-point is the mean and each error-bar the standard-deviation of 20 experiments.

While other scattering based techniques have been demonstrated for nanoparticle sizing, a majority of these techniques rely on tracking of individual particles in order to detect their mean square displacement (MSD) from which diffusion coefficients can be extracted [56, 57]. Our method, on the other hand, relies on Fourier analysis (DPS) and no tracking. This is made possible by the spatial scale information that is available from the MISS full-field phase image due to measurement of all scattering angles simultaneously. Compared to particle tracking approaches our method, thus, requires shorter acquisition times as the mean squared displacement (MSD) does not need to be mapped out for a time interval that is statistically representative of particle motion. As an example, the Nanoparticle Tracking Approach (NTA) used by Nanosight (Malvern Instruments) typically acquires images for 30-90 sec [58]. In our approach, the typical acquisition times ranged from 0.6-5 sec in total, depending on the size of the particle. Furthermore, MSD based approaches can be confounded by any translational (deterministic) motion of particles since particle tracking does not separate the deterministic from random motion in terms of spatial scales. In our DPS based approach, the diffusion coefficient D is separable from the advection velocity coefficient $\Delta v$ [Eq. (2)] because we can relate these two metrics to the spatial frequencies of the acquired complex image field [55]. Finally, another important advantage of our DPS approach is that it can be applied to continuous media, which do not exhibit discrete, trackable objects – required for particle tracking based approaches.

While traditional dynamic light scattering (DLS) does not require individual particle tracking either, most DLS measurements are at a small number of scattering angles at a time [50]. To provide spatial frequency information in the way that MISS does, DLS would require a large number of scattering angles measured simultaneously. Thus, spatial Fourier processing on these data is not possible.

### 2.4 Measurement of membrane dynamics in electrically active neurons

Next, we applied high-speed MISS imaging to biological specimens. Measurement of cell membrane displacement can reveal important information about a cell's mechanical properties including its elastic, shear and bending moduli [19, 59-61]. These mechanical properties are tightly regulated under homeostasis and deviations from physiological values may indicate an unhealthy cell [59]. The OPD measured in MISS can be used to map out fluctuations in membrane displacement as a function of time, providing a non-invasive method of investigating cell mechanics. In live cells, these fluctuations can be driven both by thermal and active processes [62]. In the context of neurons, for example, in addition to thermal fluctuations, mechanical waves travelling along the axon have also been shown to correlate with AP propagation, meaning that neuronal electrical activity also can potentially be measured through QPI [61, 63]. In either case, investigation of the cell membrane dynamics requires nanometer spatial and millisecond temporal resolution. We used the MISS microscope to measure the membrane fluctuations of primary rat cortical neurons (see Section 4: Materials and Methods for details on sample preparation). We obtained time-lapse stacks of neuron images at 833 fps before and immediately after high $K^+$ stimulation. In order to prove that the viability of neurons was unaffected by the preparation we confirmed the increase in intracellular $Ca^{+2}$ ion concentration in the neuron, due to excitation, using a fluorescence reporter (Section 4: Materials and Methods). The reporter generates a fluorescence signal with an intensity that is proportional to the $Ca^{+2}$ ion concentration. Fluorescence images of the neurons before and after chemical stimulation confirmed that the neurons were electrically active.

The results of this investigation are summarized in Fig 6. Figures 6 (a) and 6 (b) show the fluorescence images of a neuron before and after stimulation with high $K^+$. A clear increase in fluorescence was observed on stimulation for the neuron confirming AP activity. Phase images of the same neuron, before and after excitation, are shown in Fig. 6 (c) and (d). These images are representative of the 1024 frames acquired in each case for this field of view. Figures 6 (e) and (f) show the temporal standard deviation map of each OPD stack, overlaid with the respective phase images. These maps clearly show higher OPD fluctuations within the cell body compared with those seen in non-cellular areas due to noise.

The results in Fig. 6 show that there is no significant difference between the fluctuations in the pre- vs. post-excitation neurons. These findings indicate that if neuron APs generate membrane motions, they must be below our detectability limit. While experimental data quantifying AP-induced membrane motion in single mammalian neurons is lacking, theoretical models have estimated membrane displacements as small as 0.1 nm [63]. Such a displacement value translates into an even smaller number for OPD, as displacement is multiplied by the refractive index contrast between the cell and surrounding fluid, which is typically < 5%. This predicted OPD value is at the limit of our detection even for spatiotemporally averaged signals. The displacements detected in our measurements are clearly due to Brownian motion, which allows us to describe the cell membrane fluctuations using an equilibrium thermodynamic model, as follows.

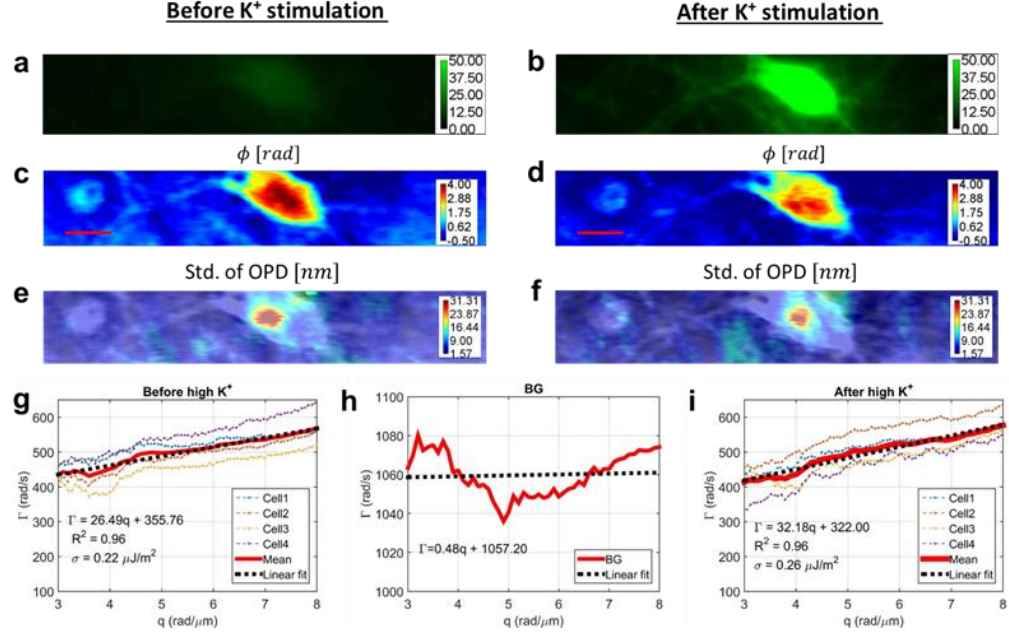

Fig. 6. Measurement of neuron membrane dynamics and surface tension σ before and after high K$^+$ stimulation. (a-b) Fluorescence images of a neuron before and after stimulation, respectively. The enhanced fluorescence signal after stimulation reports on the intracellular Ca$^{+2}$ ion concentration. (c-d) Representative phase images of the neuron. (e-f) Temporal standard deviation of the OPD time-lapse overlaid over the corresponding phase image. (g) DPS analysis on four different cells before high K$^+$ stimulation. Linear fitting of the average dispersion curve yields the mean surface tension $\sigma$ of the neuron membrane. (h) Dispersion curve for a no-sample, i.e., background (BG) region. (i) DPS analysis on the same four cells after high K$^+$ stimulation. Linear fitting of the average dispersion curve was used to extract mean surface tension $\sigma$ of the neuron membrane.

We used the nanoscale membrane fluctuations to extract the effective membrane surface tension $\sigma$ of the neuronal plasma membrane, using thermodynamic models previously published in the literature [61]. Developed originally for the erythrocyte plasma membrane, these models relate the temporal bandwidth $\Gamma$ of the fluctuations to the mechanical parameters of the underlying phospholipid bilayer/cytoskeleton. Assuming a free bilayer, the relationship between $\Gamma$ and the bending modulus $\kappa$, surface tension $\sigma$, and confinement parameter $\gamma$ of the membrane is given by the expression [61, 64]

$$\Gamma = \left[ \kappa q^3 + \sigma q + \frac{\gamma}{q} \right] \frac{1}{4\eta} \qquad (4)$$

where $\eta \approx 3\eta_{water}$ is the average dynamic viscosity of the cytoplasmic and extracellular solutions.

We extracted the dispersion relation $\Gamma(q)$ for neuron time-lapse images using DPS (see Section 4: Materials and Methods) for fitting model parameters. The results of the analysis on neuron data are illustrated in Figs. 6 (g-i). We show that the results in terms of the dispersion relation from 4 different cells yield approximately a linear behavior within the spatial frequency range $q = 3 - 8$ rad/µm. This trend indicates tension-dominated membrane fluctuations and fitting the data allows us to extract the surface tension coefficient $\sigma$ via Eq. (4). For comparison, at the same spatial scales, DPS analysis on a background image stack (no cells) showed a relatively constant relationship between $\Gamma$ and $q$. Thus, MISS imaging provides a non-invasive method of probing cell surface tension which in more traditional methods requires contact or manipulation of the cell [65-67]. To our knowledge, the measurement of mammalian neuron membrane tension has not been carried out before using microscopy methods. This tension has been measured through tethering force measurement (using optical tweezers) and micropipette aspiration [68, 69]. The reported values in those works were larger than ours, which can be understood as being due to the fact that contact methods involve large deformations in the cell membrane, which can trigger a nonlinear, higher apparent stiffness response.

## 3. Summary and Conclusions

In summary, we developed a new imaging system, capable of providing highly sensitive phase information at fast acquisition rates. The sub-nanometer sensitivity to OPD changes is due to the particular interferometric geometry used, which forces the two interfering beams to travel along a common path, thus encountering similar noise perturbations. Since we measure the phase difference between the two beams, the common noise is eliminated and, thus, the measurement is intrinsically stable. Because both beams traverse the sample, we perform optical low-pass filtering on one of the beams to convert it into a uniform reference field. The key element for accomplishing this task is a GRIN lens, which helps magnify one of the fields by a factor of 500. Thus, the camera records only the center portion of this beam, which is the uniform, DC component. The use of the microlens makes the instrument robust and easy to align. This geometry also allows for more efficient signal collection of the reference field compared with other geometries such as the DPM.

In essence, the MISS module upgrades an existing microscope and grants it full dynamic light scattering capabilities. Typically, nanoparticle sizing is accomplished either by dedicated angular scattering instruments or by instruments employing particle tracking [40, 58]. Here we showed that, exploiting the phase information, MISS can extract quantitative information from Brownian particles and size them with high precision via a much simpler dispersion relation based analysis. Since the accuracy in the size measurement is not physically limited by the numerical aperture of the objective, there is no lower bound on the particle size that MISS can characterize. We demonstrated accurate measurements down to 20 nm gold nanoparticles. Finally, we imaged nanoscale fluctuations in live neurons at thermal equilibrium and discovered the signature of a tension-dominated motion. Note that previously it has been shown in the context of red blood cells that an apparent tension-dominated motion is consistent with a geometric coupling between bending and shear modes [7]. Demonstration of this interaction in neurons is interesting and motivates future studies. We anticipate that MISS will become a broadly-adopted instrument for studying dynamics at the nanoscale and millisecond scales in both material and life sciences.

## 4. Materials and Methods

### 4.1 Calculation of the dispersion relation

Our analysis of time-resolved MISS data involves computing a dispersion relation (see Section 2: Results and Discussion) which relates the temporal bandwidth $\Gamma$ to the spatial frequency $q$ [53, 54]. Our calculation of the dispersion relation follows the dispersion-relation phase spectroscopy (DPS) method previously published in [53].

We compute this dispersion relation in the time domain to reduce the computational overhead. For a stack of time-lapse phase images, first we take the 2D Fourier transform in space for each frame. Since $\Gamma$ is the second moment of the mean-centered temporal power spectrum $P(\omega)$, it can be related to the time derivative of the measured phase $\phi(t)$ by using Parseval's theorem:

$$\Gamma = \frac{\int \omega^2 P(\omega) d\omega}{\int P(\omega) d\omega}$$

$$= \frac{\int \left|\frac{d\phi(t)}{dt}\right|^2 dt}{\int |\phi(t)|^2 dt}. \quad (5)$$

Thus, in our experiments we use the time derivative of the phase time-lapse, in spatial frequency domain, to extract $\Gamma$ for a certain spatial frequency vector $q = (q_x, q_y)$. Applying the computation for all spatial frequencies, we obtain the temporal bandwidth map $\Gamma(q_x, q_y)$. The one-dimensional profile $\Gamma(q)$ is then obtained via radially averaging $\Gamma(q_x, q_y)$, which allows for polynomial fitting (see Section 2: Results and Discussion).

### 4.2 Neuron culture and buffer solution

For neuron imaging the cell culture was prepared as follows. Primary dissociated cortical neurons were prepared from the cortex of Sprague-Dawley rat embryos. These cortical neurons were then plated on 29 mm petri dishes and pre-coated with poly-D-lysine (0.1 mg/ml; Sigma-Aldrich). Neurons (300 cells/mm$^2$) were initially incubated with a plating medium containing 86.55% MEM Eagle's with Earle's BSS (Lonza), 10% Fetal Bovine Serum (re-filtered, heat inactivated; ThermoFisher), 0.45% of 20% (wt./vol.) glucose, 1x 100 mM sodium pyruvate (100x; Sigma-Aldrich), 1x 200 mM glutamine (100x; Sigma-Aldrich), and 1x Penicillin/ Streptomycin (100x; Sigma-Aldrich). After five hours of incubation at 37°C and 5% $CO_2$, the whole media was aspirated and replaced with maintenance media. The neurons were maintained in standard maintenance media containing Neurobasal growth medium supplemented with B-27 (Invitrogen), 1% 200 mM glutamine (Invitrogen) and 1% - penicillin/streptomycin (Invitrogen) at 37 $^0$C, in the presence of 5% $CO_2$. Half the media was aspirated once a week and replaced with fresh maintenance media warmed to 37 °C. The neurons were grown in this condition for 14 days in vitro before imaging.

For imaging, the maintenance media was replaced by a modified Tyrode solution (buffer) using a formulation adopted from [70]. In order to study dynamics of electrically active neurons, we stimulated action potentials (AP) using a high $K^+$ solution. Since the transmembrane potential in a neuron is regulated by maintenance of intra- and extracellular $K^+$ ion concentration, increasing the extracellular concentration of $K^+$ ions is a convenient way of stimulating APs in neurons. We achieved stimulation by increasing the concentration of KCl in the solution from 4 mM to 45 mM with an equimolar decrease in NaCl concentration in order to maintain physiological osmolality [71]. Furthermore, the intracellular $Ca^{+2}$ concentration in the neuron was assayed using a fluorescence reporter (Fluo-4 calcium imaging kit, Sigma-Aldrich). The reporter generates a fluorescence signal with an intensity that is proportional to the $Ca^{+2}$ ion concentration. Since intracellular $Ca^{+2}$ concentration increases on AP propagation, fluorescence images of the neurons were obtained using a FITC filter cube before and after chemical stimulation to confirm neuron firing. For time-lapse MISS imaging, stacks were acquired, both before and after neuron excitation, through an automated scanning platform developed in-house [72]. Automated acquisition ensured that the time lag between the stimulus being added and the post-excitation acquisition was limited only to a few seconds. Each time-lapse stack consisted of 1024 images and the acquisition frame rate was maintained at 833 frames per second (fps) for all neuron imaging.

**Appendix A: Signal efficiency comparison between MISS and DPM**

The phase sensitivity of an off-axis interferometry system improves with the measured interferometric visibility or fringe contrast [47]. By replacing, the pinhole used in DPM with the GRIN lens (Fig. 1 in the main text) the efficiency of the reference wave signal in MISS is improved which results in an improvement in the fringe contrast and, thus, phase sensitivity.

To demonstrate this comparison, we acquired a stack of 1024 images using both DPM and MISS at 813 frames per second (fps) using a 1.2 ms exposure time. To make the comparison fair, the size of the beam entering lens $L_1$ (Fig. 1) was maintained at 4.12 mm (using an iris at the output port of the microscope) for both MISS and DPM acquisitions. A 10 µm pinhole (Edmund Optics) was used for low-pass filtering during DPM acquisition. The image size in both cases was maintained at 768x256 pixels (5 mm x 1.6 mm) which made the equivalent pinhole size (projected at the back Fourier plane of $L_1$) approx. 10 µm for the MISS system. Aside from replacing the pinhole with the GRIN lens, all other optical components were identical for the two systems.

The results of this experiment are illustrated in Fig. 7. Figs. 7 (a) and (b) show representative off-axis inteferograms obtained from the two imaging modalities. These were obtaining by temporally averaging the 1024 image stacks acquired in each case. Fig. 7 (c) shows the histograms of these two intensity images. As shown, the bimodal distribution obtained shows a greater separation between the means of the two peaks in the case of MISS, in comparison with DPM, implying greater fringe contrast. To quantify this difference, each bimodal distribution was fitted as a mixture of two unimodal normal distributions (with means $\mu_1$ and $\mu_2$ and standard deviations $\sigma_1$ and $\sigma_2$) using a maximum likelihood estimator (MLE) in MATLAB [73]. The results of the fit illustrate that MISS provides approximately a factor of 2 better fringe contrast than DPM, as evidenced by the difference in means ($\mu_2 - \mu_1$) for the two distributions. Since the sample wave (first diffraction order) is identical for both modalities, the MISS microscope provides greater efficiency for the reference order signal, improving fringe contrast and, thus, phase sensitivity.

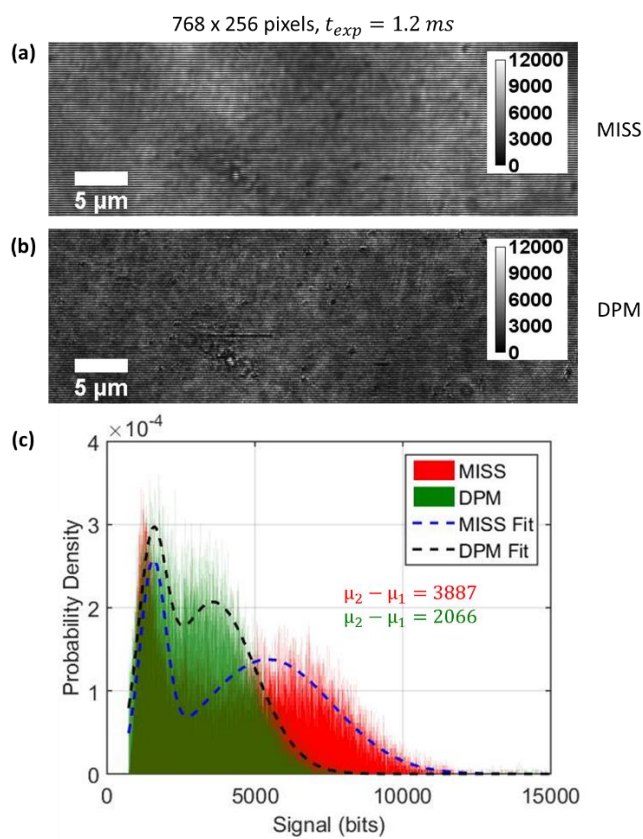

Fig. 7. (a) and (b) Average raw intensity images obtained for MISS and DPM, respectively, by temporally averaging 1024 frames. (c) Histograms of the intensity images in (a) and (b). Fitting a mixture of two normal distributions to the bimodal histograms show a greater separation between the peaks in MISS than in DPM.

**Funding:**

This work was supported by the National Science Foundation (NSF) (CBET-1040461 MRI, CBET-0939511 STC, DBI 1450962 EAGER and IIP-1353368).

**Acknowledgments:**

We would like to thank Daniel Fernandes for his help in building of the optical setup and Sung-soo Jang and Dr. Hee Jung Chung for their inputs regarding the neuron cell culture.

**Author contributions:**

H. Majeed and G. Popescu conceived the idea of MISS. H. Majeed was involved in the design, construction and characterization of the optical system as well as in the live neuron imaging. L. Ma gathered data and conducted analysis for the nanoparticle sizing experiments. Y. J. Lee cultured the live neurons and helped with neuron imaging. M. Kandel built the software platform used for the neuron imaging. E. Min and W. Jung helped with optical system design. C. Best-Popescu led the neuron imaging study. G. Popescu supervised the project.

**Financial Interest Disclosure:**

Gabriel Popescu has financial interest in Phi Optics, Inc., a company that develops quantitative phase imaging technologies.